\documentclass[a4paper]{article}

\usepackage{INTERSPEECH2020}
\usepackage{amsmath}
\usepackage{newtxmath}
\usepackage{hyperref}
\usepackage{xcolor}
\usepackage{cite}

\makeatletter
\newcommand{\thickhline}{%
	\noalign {\ifnum 0=`}\fi \hrule height 1.0pt
	\futurelet \reserved@a \@xhline
}

\title{VQMIVC: Vector Quantization and Mutual Information-Based Unsupervised Speech Representation Disentanglement for One-shot Voice Conversion}
\name{Disong Wang$^{1*}$\thanks{*Work done during internship at Huawei Noah’s Ark Lab}, Liqun Deng$^2$, Yu Ting Yeung$^2$, Xiao Chen$^2$, Xunying Liu$^1$, Helen Meng$^1$}
\address{
  $^1$The Chinese University of Hong Kong, Hong Kong SAR, China\\
  $^2$Huawei Noah’s Ark Lab}
\email{\footnotesize \{dswang,xyliu,hmmeng\}@se.cuhk.edu.hk, \{dengliqun.deng,yeung.yu.ting,chen.xiao2\}@huawei.com}

\begin{document}

\maketitle
\begin{abstract}
  One-shot voice conversion (VC), which performs conversion across arbitrary speakers with only a single target-speaker utterance for reference, can be effectively achieved by speech representation disentanglement. Existing work generally ignores the correlation between different speech representations during training, which causes leakage of content information into the speaker representation and thus degrades VC performance. To alleviate this issue, we employ vector quantization (VQ) for content encoding and introduce mutual information (MI) as the correlation metric during training, to achieve proper disentanglement of content, speaker and pitch representations, by reducing their inter-dependencies in an unsupervised manner. Experimental results reflect the superiority of the proposed method in learning effective disentangled speech representations for retaining source linguistic content and intonation variations, while capturing target speaker characteristics. In doing so, the proposed approach achieves higher speech naturalness and speaker similarity than current state-of-the-art one-shot VC systems. Our code, pre-trained models and demo are available at \color{blue}\url{https://github.com/Wendison/VQMIVC}.
\end{abstract}

\section{Introduction}

Voice conversion (VC) is a technique used to modify para-linguistic factors of an utterance from a source speaker to sound like a target speaker. Para-linguistic factors include speaker identity \cite{mohammadi2017overview}, prosody \cite{rentzos2003transformation} and accent \cite{oyamada2017non}, etc. In this paper, we focus on the conversion of speaker identity across arbitrary speakers under a one-shot scenario \cite{liu2018voice,lu2019one}, i.e., given only one target speaker's utterance for reference. 

Previous work that use methods based on speech representation disentanglement (SRD) \cite{qian2019autovc,chou2019one,wu2020vqvc+} attempted to address one-shot VC by decomposing the speech into speaker and content representations, and then the speaker identity can be converted by changing the source speaker's representation to that of the target speaker. However, it is difficult to measure the degree of SRD. Besides, previous approaches generally do not impose correlation constraints between speaker and content representations during training, which results in leakage of content information into the speaker representation, leading to VC performance degradation. To alleviate these issues, this paper proposes the vector quantization and mutual information-based VC (VQMIVC) approach, where mutual information (MI) measures the dependencies between different representations and can be effectively integrated into the training process to achieve SRD in an \textit{unsupervised manner}. Specifically, we first decompose an utterance into three factors: content, speaker and pitch, and then propose a VC system consisting of four components: (1) A content encoder using vector quantization with contrastive predictive coding (VQCPC) \cite{van2020vector,baevski2019vq} to extract frame-level content representations from acoustic features; (2) A speaker encoder that takes in acoustic features to generate a single fixed-dimensional vector as the speaker representation; (3) A pitch extractor that is used to compute normalized fundamental frequency ($\textit{F}_0$) at the utterance level as the pitch representation; and (4) A decoder that maps content, speaker and pitch representations to acoustic features. During training, the VC system is optimized by minimizing VQCPC, reconstruction and MI losses. VQCPC aims to explore local structures of speech, and MI reduces the inter-dependencies of different speech representations. During inference, one-shot VC is achieved by only replacing the source speaker representation with the target speaker representation derived from a single target utterance. The main contribution of this work lies in applying the combination of VQCPC and MI to achieve SRD, without any requirements of supervision information such as text transcriptions or speaker labels. Extensive experiments have been conducted to thoroughly analyze the importance of MI, where information leakage issues can be significantly alleviated for enhanced SRD.

% The contributions of this paper are two folds: (1) The proposed approach employs VQ and MI simultaneously to achieve SRD in a fully unsupervised manner; (2) 

\section{Related work}
VC performance is critically dependent on the availability of the target speaker’s voice data for training \cite{toda2007voice,helander2010voice,erro2009inca,sun2016phonetic,hsu2017voice,kaneko2017parallel,kameoka2018stargan}. Hence, the challenge of one-shot VC is in performing conversion across arbitrary speakers that may be unseen during training, and with only one single target-speaker utterance for reference. Previous approaches for one-shot VC are based on SRD, which aims to separate speaker information from spoken content as far as possible. Related work include: tunable information-constraining bottleneck \cite{qian2019autovc,qian2020f0,qian2020unsupervised}, instance normalization techniques \cite{chou2019one,chen2021again} and vector quantization (VQ) \cite{wu2020vqvc+,chorowski2019unsupervised}. We adopt VQCPC \cite{van2020vector,baevski2019vq}, which is an improved version of VQ, to extract accurate content representations. Without explicit constraints between different speech representations, information leakage tends to occur, which degrades VC performance. We draw inspirations from information theory \cite{gierlichs2008mutual} in using MI as a regularizer to constrain the dependency between variables. As MI computation is challenging for variables with unknown distribution, various methods have been explored to estimate MI lower bound \cite{nguyen2010estimating,gutmann2010noise,belghazi2018mutual} in SRD-based speech tasks \cite{ravanelli2019learning,kwon2020intra,hu2020unsupervised} . To guarantee the reduction of MI values, we propose to use variational contrastive log-ratio upper bound (vCLUB) \cite{cheng2020club}. While a recent effort \cite{yuan2021improving} employs MI for VC by using speaker labels as the supervision for learning speaker representations, our proposed approach differs from \cite{yuan2021improving} in terms of the combination of VQCPC and MI for fully unsupervised training, and the incorporation of pitch representations to maintain source intonation variations.

\begin{figure}[t]
  \centering
  \centerline{\includegraphics[width=7cm]{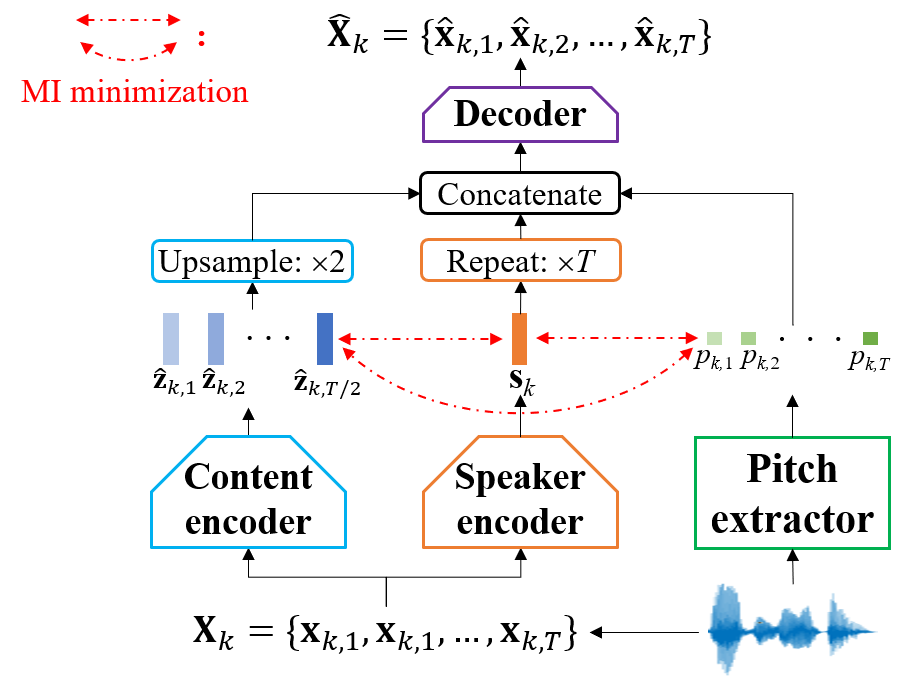}}
   \vspace{-1em}
  \caption{Diagram of the proposed VQMIVC system.}\label{arc}
  \vspace{-1.5em}
\end{figure}

\section{Proposed approach}

This section first describes the system architecture of the VQMIVC approach, then elaborates on the integration of MI minimization into the training process, and finally shows how one-shot VC is achieved. 

\subsection{Architecture of the VQMIVC system}
As shown in Figure \ref{arc}, the proposed VQMIVC system includes four modules: content encoder, speaker encoder, pitch extractor and decoder. The first three modules respectively extract content, speaker and pitch representations from the input voice; and the fourth module, the decoder, maps these representations back to acoustic features. Assuming that there are \textit{K} utterances, we use mel-spectrograms as acoustic features and randomly select \textit{T} frames from each utterance for training. The $\textit{k}^{th}$ mel-spectrogram is denoted as $\textbf{X}_k=\{\textbf{x}_{k,1},\textbf{x}_{k,2},...,\textbf{x}_{k,T}\}$.

\vspace{0.1em}

\begin{figure}[t]
  \centering
  \centerline{\includegraphics[width=5.5cm]{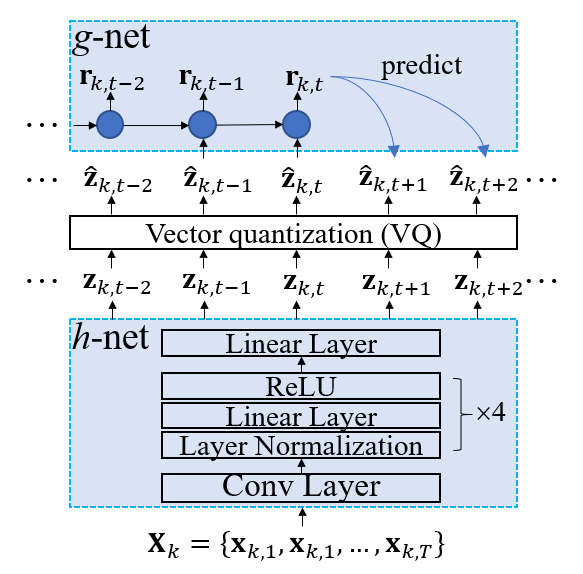}}
  \vspace{-1em}
  \caption{Details of the VQCPC based content encoder.}\label{ce}
  \vspace{-1.5em}
\end{figure}

\textbf{Content encoder} ${\theta}_c$: The content encoder strives to extract linguistic content information from $\textbf{X}_k$ by using VQCPC as shown in Figure \ref{ce}, which contains two networks \textit{h}-net: $\textbf{X}_k$ $\rightarrow$ $\textbf{Z}_k$ and \textit{g}-net: $\hat{\textbf{Z}}_k$ $\rightarrow$ $\textbf{R}_k$, and VQ operation \textit{q}: $\textbf{Z}_k$ $\rightarrow$ $\hat{\textbf{Z}}_k$. \textit{h}-net takes in $\textbf{X}_k$ to derive a sequence of dense features $\textbf{Z}_k=\{\textbf{z}_{k,1},\textbf{z}_{k,2},...,\textbf{z}_{k,T/2}\}$, where the length is reduced from \textit{T} to \textit{T/2}. Then the quantizer \textit{q} discretizes $\textbf{Z}_k$ with a trainable codebook \textit{B} into $\hat{\textbf{Z}}_k$=\{$\hat{\textbf{z}}_{k,1}$,$\hat{\textbf{z}}_{k,2}$,...,$\hat{\textbf{z}}_{k,T/2}$\}, where $\hat{\textbf{z}}_{k,t}$ $\in \textit{B}$ is the vector closest to $\textbf{z}_{k,t}$. VQ imposes an information bottleneck to remove non-essential details in $\textbf{Z}_k$, making $\hat{\textbf{Z}}_k$ to be related with underlying linguistic information. Then the content encoder ${\theta}_c$ is trained by minimizing the VQ loss \cite{van2020vector}:
\begin{footnotesize}
\begin{equation}
{{L}_{VQ}}=\frac{2}{KT}\sum\limits_{k=1}^{K}{\sum\limits_{t=1}^{T/2} \left\| {{\textbf{z}}_{k,t}}-sg({{{\hat{\textbf{z}}}}_{k,t}}) \right\|_{2}^{2}} \label{vq-loss}
\end{equation}
\end{footnotesize}where \textit{sg}(·) denotes the stop-gradient operator. To further encourage $\hat{\textbf{Z}}_k$ to capture local structures, contrastive predictive coding (CPC) is employed by adding RNN based \textit{g}-net taking in $\hat{\textbf{Z}}_k$ to obtain aggregation $\textbf{R}_k=\{\textbf{r}_{k,1},\textbf{r}_{k,2},...,\textbf{r}_{k,T/2}\}$. Given $\textbf{r}_{k,t}$, the model is trained to distinguish a positive sample $\hat{\textbf{z}}_{k,t+m}$ that is \textit{m} steps in the future from negative samples drawn from the set ${\Omega}_{k,t,m}$ by minimizing the InfoNCE loss \cite{oord2018representation}:
\begin{footnotesize}
\begin{equation}
{{L}_{CPC}}=-\frac{1}{KT'M}\!\sum\limits_{k=1}^{K}{\sum\limits_{t=1}^{T'}{\sum\limits_{m=1}^{M}{\!\log\! \left[\! \frac{\exp (\hat{\textbf{z}}_{k,t+m}^{T}{{\mathbf{W}}_{m}}{{\textbf{r}}_{k,t}})}{\sum\nolimits_{\tilde{\textbf{z}}\in {{\Omega }_{k,t,m}}}{\exp ({{{\tilde{\textbf{z}}}}^{T}}{{\mathbf{W}}_{m}}{{\textbf{r}}_{k,t}})}} \!\right]}}} \label{cpc-loss} 
\end{equation}
\end{footnotesize}where ${T'}=T/2-M$, $\textbf{W}_m$ (\textit{m}=1,2,...,\textit{M}) is trainable projection matrix. By predicting future samples with probabilistic contrastive loss (\ref{cpc-loss}), local features (e.g., phonemes) spanning many time steps are encoded into $\hat{\textbf{Z}}_k$=\textit{f}($\textbf{X}_k$;${\theta}_c$), which is the content representation used to accurately reconstruct the linguistic content. During training, the negative samples set ${\Omega}_{k,t,m}$ is formed by randomly selecting samples from the current utterance.

% \vspace{-0.5em}

\textbf{Speaker encoder} ${\theta}_s$: The speaker encoder takes in $\textbf{X}_k$ to generate a vector $\textbf{s}_k$=\textit{f}($\textbf{X}_k$;${\theta}_s$), which is used as the speaker representation. $\textbf{s}_k$ captures global speech characteristics to control the speaker identity of the generated speech. 

\textbf{Pitch extractor}: The pitch representation is expected to contain intonation variations but exclude content and speaker information, so we extract $\textit{F}_0$ from the waveform and perform z-normalization for each utterance independently. In our experiments, we adopt log-normalized $\textit{F}_0$ (log-$\textit{F}_0$) as $\textbf{p}_k$=($\textit{p}_{k,1}$, $\textit{p}_{k,2}$, ..., $\textit{p}_{k,T}$), which is speaker-independent so that speaker encoder is forced to provide the speaker information, e.g., vocal ranges.

\textbf{Decoder} ${\theta}_d$: The decoder is used to map the content, speaker and pitch representations to mel-spectrograms. Linear interpolation based upsampling ($\times$2) and repetition ($\times$\textit{T}) are performed on $\hat{\textbf{Z}}_k$ and $\textbf{s}_k$ respectively to align with $\textbf{p}_k$ as inputs of decoder to generate mel-spectrograms $\hat{\textbf{X}}_k$=\{$\hat{\textbf{x}}_{k,1}$,$\hat{\textbf{x}}_{k,2}$,...,$\hat{\textbf{x}}_{k,T}$\}. Decoder is jointly trained with content and speaker encoders by minimizing a reconstruction loss:
\begin{footnotesize}
\begin{equation}
{{L}_{REC}}=\frac{1}{KT}\sum\limits_{k=1}^{K}{\sum\limits_{t=1}^{T}{\left[{{\left\| {{{\hat{\textbf{x}}}}_{t}}-{{\textbf{x}}_{t}} \right\|}_{1}}+{{\left\| {{{\hat{\textbf{x}}}}_{t}}-{{\textbf{x}}_{t}} \right\|}_{2}}\right]}} \label{rec-loss}
\vspace{-0.5em}
\end{equation}
\end{footnotesize}

\vspace{-1em}
\subsection{MI minimization integrated into VQMIVC training}
Given the random variables \textbf{u} and \textbf{v}, the MI is Kullback-Leibler (KL) divergence between their joint and marginal distributions as $\mathbf{I}(\mathbf{u},\mathbf{v})={{D}_{KL}}(P(\mathbf{u},\mathbf{v});P(\mathbf{u})P(\mathbf{v}))$. We adopt vCLUB \cite{cheng2020club} to compute the upper bound of MI as:
\begin{footnotesize}
\begin{equation}
    {\textbf{I}}\!\left( \textbf{u},\textbf{v} \right)\!=\!{{E}_{P(\textbf{u},\textbf{v})}}\!\left[\! \log {{Q}_{{{\theta }_{\textbf{u},\textbf{v}}}}}\!\left( \textbf{u}|\textbf{v} \right)\! \right]\!-\!{{E}_{P(\textbf{u})}}{{E}_{P(\textbf{v})}}\!\left[\! \log {{Q}_{{{\theta }_{\textbf{u},\textbf{v}}}}}\!\left( \textbf{u}|\textbf{v} \right)\! \right] \label{club}
\end{equation}
\end{footnotesize}where \textbf{u},\textbf{v} $\in$ \{$\hat{\textbf{Z}}$, \textbf{s}, \textbf{p}\}, $\hat{\textbf{Z}}$, \textbf{s} and \textbf{p} are content, speaker and pitch representations respectively, ${{Q}_{{{\theta }_{\textbf{u},\textbf{v}}}}}(\textbf{u}|\textbf{v})$ is the variational approximation of ground-truth posterior of \textbf{u} given \textbf{v} and can be parameterized by a network ${{\theta }_{\textbf{u},\textbf{v}}}$. The unbiased estimation for vCLUB between different speech representations is given by:
\begin{footnotesize}
\begin{equation}
    \!{{\hat{\textbf{I}}}({\hat{\textbf{Z}},\textbf{s}})}\!=\!\frac{2}{{{K}^{2}}T}\sum\limits_{k=1}^{K}{\sum\limits_{l=1}^{K}{\sum\limits_{t=1}^{T/2}{\!\left[\! \log\!{{Q}_{{{\theta }_{\hat{\textbf{Z}},\textbf{s}}}}}\!\left( {{{\hat{\textbf{z}}}}_{k,t}}|{{\textbf{s}}_{k}}\! \right)\!-\!\log\! {{Q}_{{{\theta }_{\hat{\textbf{Z}},\textbf{s}}}}}\!\left( {{{\hat{\textbf{z}}}}_{l,t}}|{{\textbf{s}}_{k}}\! \right)\! \right]}}} \label{mi-1}
\end{equation}
\end{footnotesize}
\begin{footnotesize}
\begin{equation}
    \!{{\hat{\textbf{I}}}({{\textbf{p}},\textbf{s}})}\!=\!\frac{1}{{{K}^{2}}T}\sum\limits_{k=1}^{K}{\sum\limits_{l=1}^{K}{\sum\limits_{t=1}^{T}{\left[\! \log\!{{Q}_{{{\theta }_{{\textbf{p}},\textbf{s}}}}}\!\left( {{{{\textbf{p}}}}_{k,t}}|{{\textbf{s}}_{k}}\! \right)\!-\!\log\! {{Q}_{{{\theta }_{{\textbf{p}},\textbf{s}}}}}\!\left( {{{{\textbf{p}}}}_{l,t}}|{{\textbf{s}}_{k}}\! \right)\! \right]}}} \label{mi-2}
\end{equation}
\end{footnotesize}
\begin{footnotesize}
\begin{equation}
    \!\!\hat{\textbf{I}}(\!\hat{\textbf{Z}},\!\textbf{p}\!)\!=\!\frac{2}{{{K}^{2}}T}\!\sum\limits_{k=1}^{K}{\!\sum\limits_{l=1}^{K}{\!\sum\limits_{t=1}^{T/2}{\!\left[\! \log\! {{Q}_{{{\theta }_{\hat{\textbf{Z}},\!\textbf{p}}}}}\!\!\left( {{{\hat{\textbf{z}}}}_{k,t}}\!|{{{\hat{\textbf{p}}}}_{k,t}} \right)\!-\!\log\! {{Q}_{{{\theta }_{\hat{\textbf{Z}},\!\textbf{p}}}}}\!\!\left( {{{\hat{\textbf{z}}}}_{l,t}}\!|{{{\hat{\textbf{p}}}}_{k,t}} \!\right)\! \right]}}} \label{mi-3}
\end{equation}
\end{footnotesize}where $\hat{\textbf{p}}_{k,t}=(\textbf{p}_{k,{2t-1}}+\textbf{p}_{k,{2t}})/2$. With a good variational approximation, (\ref{club}) provides a reliable MI upper bound. Therefore, we can decrease the correlation among different speech representations by minimizing (\ref{mi-1})-(\ref{mi-3}), and the total MI loss is:
\vspace{-0.2em}
\begin{footnotesize}
\begin{equation}
    % {{L}_{MI}}=\hat{\textbf{I}}\left( \hat{\textbf{Z}},\mathbf{s} \right)\text{+}\hat{\textbf{I}}\left( \hat{\textbf{Z}},\mathbf{p} \right)\text{+}\hat{\textbf{I}}\left( \mathbf{p},\mathbf{s} \right) \label{mi-loss}
    {{L}_{MI}}=\hat{\textbf{I}}(\hat{\textbf{Z}},\mathbf{s})\text{+}\hat{\textbf{I}}(\hat{\textbf{Z}},\mathbf{p})\text{+}\hat{\textbf{I}}(\mathbf{p},\mathbf{s}) \label{mi-loss}
\end{equation}
\end{footnotesize}During training, variational approximation networks and VC network are optimized alternatively. The variational approximation networks are trained to maximize the log-likelihood:
\vspace{-0.2em}
\begin{footnotesize}
\begin{equation}
\begin{matrix}
   {{L}_{\textbf{u},\textbf{v}}}=\log {{Q}_{{{\theta }_{\textbf{u},\textbf{v}}}}}\left( \textbf{u}|\textbf{v} \right), & \mathbf{u},\mathbf{v}\in \{\hat{\textbf{Z}},\mathbf{s},\mathbf{p}\}  \\
\end{matrix} \label{ll-loss}
\end{equation}
\end{footnotesize}while the VC network is trained to minimize VC loss:
\vspace{-0.1em}
\begin{footnotesize}
\begin{equation}
{{L}_{VC}}={{L}_{VQ}}+{{L}_{CPC}}+{{L}_{REC}}+{\lambda}_{MI}{{L}_{MI}}
\label{vc-loss}
\end{equation}
\end{footnotesize}where ${\lambda}_{MI}$ is a constant weight to control how MI loss enhances the disentanglement. The final training process is summarized in Algorithm 1. We note that no text transcriptions or speaker labels are used during training, so the proposed approach achieves disentanglement in a fully unsupervised way.

\begin{table}[t]
  %\caption{Pseudocode for the proposed VQMIVC training}
  \label{alo}
  \centering
  \begin{tabular}{ll}
    \toprule[1.0pt]
    \textbf{Algorithm 1.} Training process of the proposed VQMIVC  
    \\
    \midrule
    \textbf{Input:} mel-spectrograms \{$\textbf{X}_k$\}, normalized log-$\textit{F}_0$ \{$\textbf{p}_k$\},  
    \\
    learning rate $\alpha$ and $\beta$
    \\
    1. \textbf{for} each training iteration \textbf{do}
    \\
    2. \quad $\hat{\textbf{Z}}_k$ $\leftarrow$ \textit{f}($\textbf{X}_k$;${\theta}_c$), $\textbf{s}_k$ $\leftarrow$ \textit{f}($\textbf{X}_k$;${\theta}_s$)
    \\
    3. \quad Calculate log-likelihood ${{L}_{\textbf{u},\textbf{v}}}$ (\ref{ll-loss}), then update:
    \\
    4. \qquad \quad ${{\theta}_{\textbf{u},\textbf{v}}}\leftarrow {{\theta}_{\textbf{u},\textbf{v}}}+\alpha {\nabla}_{{\theta}_{\textbf{u},\textbf{v}}} {{L}_{\textbf{u},\textbf{v}}}$, \; \textbf{u},\textbf{v} $\in$ \{$\hat{\textbf{Z}}$, \textbf{s}, \textbf{p}\}
    \\
    5. \quad Calculate VC loss  $\textit{L}_{VC}$ (\ref{vc-loss}), then update:
    \\
    \qquad \qquad $\theta \leftarrow \theta -\beta {{\nabla }_{\theta }}{{L}_{VC}}$, \; $\theta \in \{{{\theta }_{c}},{{\theta }_{s}},{{\theta }_{d}}\}$
    \\
    6. \textbf{end for}
    \\
    7. \textbf{return} ${{\theta }_{c}}$, ${{\theta }_{s}}$, ${{\theta }_{d}}$
    \\
    \bottomrule[1.0pt]
  \end{tabular}
  \vspace{-1.3em}
\end{table}

\subsection{One-shot VC}
During conversion, the content and pitch representations are first extracted from source speaker's utterance $\textbf{X}_\textit{src}$ as $\hat{\textbf{Z}}_{src}=\textit{f}({\textbf{X}_\textit{src}};{{\theta }_{c}})$ and $\textbf{p}_{src}$ respectively, the speaker representation is extracted from only one target speaker's utterance $\textbf{X}_\textit{tgt}$ as $\textbf{s}_{tgt}=\textit{f}({\textbf{X}_{tgt}};{{\theta }_{s}})$, then the decoder generates the converted mel-spectrograms as \textit{f}$(\hat{\textbf{Z}}_{src},\textbf{s}_{tgt},\textbf{p}_{src};{\theta}_d)$.

\section{Experiments}

\subsection{Experimental setup}
All experiments are conducted on the VCTK corpus \cite{veaux2016superseded} with 110 English speakers, which are randomly split into 90 and 20 speakers as training and testing sets respectively. The testing speakers are treated as unseen speakers that are used to perform one-shot VC. For acoustic features extraction, all audio recordings are downsampled to 16kHz, 80-dim mel-spectrograms and $\textit{F}_0$ are both calculated with 25ms Hanning window, 10ms frame shift and 400-point fast Fourier transform.

The proposed VC network consists of the content encoder, speaker encoder and decoder. The content encoder contains a \textit{h}-net, a quantizer \textit{q} and a \textit{g}-net. The \textit{h}-net is composed of a convolutional layer with stride of 2, four blocks with layer normalization, 512-dim linear layer and ReLU activation function for each block. The quantizer contains a codebook with 512 64-dim learnable vectors. The \textit{g}-net is a 256-dim uni-directional RNN layer. For CPC, the future prediction step \textit{M} is 6 and the number of negative samples $|{\Omega}_{k,t,m}|$ is 10. The speaker encoder follows \cite{chou2019one}, which contains 8 ConvBank layers to encode the long-term information, 12 convolutional layers with 1 average-pooling layer, and 4 linear layers to derive the 256-dim speaker representation. The decoder follows \cite{qian2019autovc} with a 1024-dim LSTM layer, three convolutional layers, two 1024-dim LSTM layers and a 80-dim linear layer. Besides, a 5-layer convolutional based Postnet is added to refine predicted mel-spectrograms, which are converted to waveform by Parallel WaveGAN vocoder \cite{yamamoto2020parallel} that is trained by VCTK corpus. The variational approximation ${{Q}_{{{\theta }_{\textbf{u},\textbf{v}}}}}(\textbf{u}|\textbf{v})$ for all MI is parameterized in Gaussian distribution as ${{Q}_{{{\theta }_{\textbf{u},\textbf{v}}}}}(\textbf{u}|\textbf{v})=\mathcal{N} (\textbf{u}|\mu (\textbf{v}),diag({{\sigma }^{2}}(\textbf{v})))$ with mean $\mu (\textbf{v})$ and variance ${\sigma }^{2}(\textbf{v})$ inferred by a two-way fully-connected network ${{\theta }_{\textbf{u},\textbf{v}}}$ that is composed of four 256-dim hidden layers. The VC network is trained using Adam optimizer \cite{kingma2014adam} with 15-epoch warmup increasing the learning rate from 1e-6 to 1e-3, which is halved every 100 epochs after 200 epochs until 500 epochs in total. Batch size is 256 and 128 frames are randomly selected from each utterance for training per iteration. Variational approximation networks are also trained with the Adam optimizer with a learning rate of 3e-4. We compare our proposed VQMIVC method with AutoVC \cite{qian2019autovc}, AdaIN-VC \cite{chou2019one} and VQVC+ \cite{wu2020vqvc+}, which are among the state-of-the-art one-shot VC methods. 

% \footnote{\url{https://github.com/auspicious3000/autovc}}
% \footnote{\url{https://github.com/jjery2243542/adaptive_voice_conversion}}
% \footnote{\url{https://github.com/ericwudayi/SkipVQVC}}
% Readers are encouraged to listen to our audio samples\footnote{Audio samples: \url{https://wendison.github.io/VQMIVC-demo/}}.

\begin{table}[t]
  \caption{MI among content, speaker and pitch representations, where MI is estimated on all testing speakers for 10 rounds, and mean $\pm$ standard-variance are reported.}
  \label{mi-est}
  \vspace{-1em}
  \centering
  \scalebox{0.95}{\begin{tabular}{c|c|c|c}
    \toprule[1.0pt]
    ${\lambda}_{MI}$ & $\hat{\textbf{I}}(\hat{\textbf{Z}},\mathbf{s})$ & $\hat{\textbf{I}}(\hat{\textbf{Z}},\mathbf{p})$ & $\hat{\textbf{I}}(\mathbf{p},\mathbf{s})$              \\
    \midrule
    0 & 24.65 $\pm$ 1.79 & 29.47 $\pm$ 0.31 & 0.12 $\pm$ 0.01   \\
    1e-3 & 8.55 $\pm$ 1.26 & 2.59 $\pm$ 0.03 & 0.06 $\pm$ 0.01   \\
    1e-2 & 8.00 $\pm$ 1.07 & 0.30 $\pm$ 0.01 & \textbf{0.02 $\pm$ 0.01}   \\
    1e-1 & \textbf{5.52 $\pm$ 0.58} & \textbf{0.09 $\pm$ 0.01} & 0.03 $\pm$ 0.01   \\
    \bottomrule[1.0pt]
  \end{tabular}}
  \vspace{-0.6em}
\end{table}

\begin{table}[t]
  \caption{CER/WER for generated speech using speaker representations from Same and Mixed utterances by varying ${\lambda}_{MI}$.}
  \label{con-to-spk}
  \vspace{-1em}
  \centering
  \scalebox{0.95}{\begin{tabular}{c|c|c}
    \toprule[1.0pt]
    % \thickhline
    ${\lambda}_{MI}$ & \textit{Same}: CER / WER & \textit{Mixed}: CER(${\triangle}_C$) / WER(${\triangle}_W$) \\
    \midrule
    0 & 23.4\% / 32.1\% & 35.9\%(12.5\%) / 59.9\%(27.8\%) \\
    1e-3 & 12.8\% / 25.6\% & 13.9\%(1.1\%) / 27.7\%(2.1\%) \\
    1e-2 & \textbf{12.3\%} / \textbf{24.9\%} & \textbf{12.9\%}(0.6\%) / \textbf{25.9\%}(1.0\%) \\
    1e-1 & 12.7\% / 25.6\% & \textbf{12.9\%}(\textbf{0.2\%}) / \textbf{25.9\%}(\textbf{0.3\%}) \\
    \bottomrule[1.0pt]
  \end{tabular}}
  \vspace{-1.5em}
\end{table}

\subsection{Experimental results and analysis}
\subsubsection{Speech representation disentanglement performance}
In the VC loss (\ref{vc-loss}), ${\lambda}_{MI}$ determines the capacity of MI to enable SRD, we first vary ${\lambda}_{MI}$ to evaluate disentanglement degrees between different speech representations extracted from all testing utterances  by computing vCLUB as shown in Table \ref{mi-est}. We can see that when ${\lambda}_{MI}$ increases, MI tends to decrease to reduce the correlation among different speech representations. 

To measure how much content information is entangled with speaker representation, we adopt two ways to generate the speech, i.e., (1) \textit{Same}, i.e., the content, speaker and pitch representations of the same utterance are used to generate the speech; (2) \textit{Mixed}, i.e., the content and pitch representations of one utterance and speaker representation of another utterance are used to generate the speech, both utterances belong to the same speaker. Then an automatic speech recognition (ASR) system is used to obtain character/word error rate (CER/WER) of the generated speech. The increased CER and WER from `\textit{Same}' to `\textit{Mixed}' are denoted as ${\triangle}_C$ and ${\triangle}_W$ respectively. As the only difference in inputs for speech generation is that of speaker representation, we can conclude that larger values of ${\triangle}_C$ and ${\triangle}_W$ reflect that more content information is leaked to speaker representation. All testing speakers are used for speech generation, and the publicly released Jasper-based ASR system \cite{li2019jasper} is used. The results are shown in Table \ref{con-to-spk}, we can see that when MI is not used (${\lambda}_{MI}$=0), the generated speech is severely contaminated by the undesired content information that resides in the speaker representations as indicated by the largest values of ${\triangle}_C$ and ${\triangle}_W$. However, when MI is used (${\lambda}_{MI}${\textgreater}0), significant reductions of ${\triangle}_C$ and ${\triangle}_W$ can be obtained. As ${\lambda}_{MI}$ increases, both ${\triangle}_C$ and ${\triangle}_W$ decrease, showing that higher ${\lambda}_{MI}$ can, to a larger degree, alleviate leakage of content information into the speaker representation.

In addition, we design two speaker classifiers, taking $\hat{\textbf{Z}}$ and \textbf{s} as inputs respectively; and one predictor, taking in $\hat{\textbf{Z}}$ to infer \textbf{p}. The classifiers and predictor are all 4-layer fully-connected network with 256-dim hidden size. Higher speaker classification accuracy denotes more speaker information in $\hat{\textbf{Z}}$ or \textbf{s}, while higher prediction loss (mean square error) for \textbf{p} denotes less pitch information in $\hat{\textbf{Z}}$. The results are shown in Table \ref{acc-loss}. We can observe that $\hat{\textbf{Z}}$ contains less speaker and pitch information when ${\lambda}_{MI}$ increases to achieve lower accuracy and higher pitch loss. Speaker classification accuracy on \textbf{s} is high for all ${\lambda}_{MI}$, while the accuracy decreases when ${\lambda}_{MI}$ increases, showing that \textbf{s} contains abundant speaker information, but higher ${\lambda}_{MI}$ tends to make \textbf{s} lose speaker information. To ensure proper disentanglement, we set ${\lambda}_{MI}$ to 1e-2 for the following experiments.  

\begin{table}[t]
  \caption{Speaker classification accuracy on $\hat{\textbf{Z}}$ (content) and \textbf{s} (speaker), and prediction loss for \textbf{p} (pitch) inferred by $\hat{\textbf{Z}}$.}
  \label{acc-loss}
  \vspace{-1em}
  \centering
  \scalebox{0.95}{\begin{tabular}{c|c|c|c}
    \toprule[1.0pt]
    ${\lambda}_{MI}$ & $\hat{\textbf{Z}}$-accuracy & \textbf{s}-accuracy & \textbf{p}-loss  \\
    \midrule
    0 & 9.7\% & \textbf{100\%} & 0.279 \\
    1e-3 & 9.5\% & 99.7\% & 0.284 \\
    1e-2 & 9.4\% & 99.5\% & 0.287 \\
    1e-1 & \textbf{9.3\%} & 98.1\% & \textbf{0.289} \\
    \bottomrule[1.0pt]
  \end{tabular}}
  \vspace{-0.6em}
\end{table}

\begin{table}[t]
  \caption{ASR and $\textit{F}_0$-PCC results for one-shot VC.}
  \label{asr-pcc}
  \vspace{-1em}
  \centering
  \scalebox{0.95}{\begin{tabular}{c|cc|c}
    \toprule[1.0pt]
    \textbf{Methods} & CER & WER & $\textit{F}_0$-PCC               \\
    \midrule
    Source (Oracle) & 3.5\% & 9.0\% & 1.0 \\
    % \midrule
    AutoVC & 15.7\% & 30.5\% & 0.455 \\
    AdaIN-VC & 27.1\% & 47.1\% & 0.346 \\
    VQVC+ & 35.5\% & 59.5\% & 0.237 \\
    \midrule
    VQMIVC (proposed) & \textbf{14.9\%} & \textbf{29.3\%} & \textbf{0.781} \\
    w/o MI (proposed) & 38.0\% & 62.1\% & \textbf{0.781} \\
    \bottomrule[1.0pt]
  \end{tabular}}
  \vspace{-1.7em}
\end{table}

\subsubsection{Content preservation and \texorpdfstring{$\textit{F}_0$}{Lg} variation consistency}
To evaluate whether the converted voice maintains linguistic content and intonation variations of the source voice, we test the CER/WER of the converted speech and calculate the Pearson correlation coefficient (PCC) \cite{benesty2009pearson} between $\textit{F}_0$ of source and converted voice. PCC ranges from -1 to 1 and can be effectively used to measure the correlation between two variables, where a higher $\textit{F}_0$-PCC denotes that the converted voice has higher $\textit{F}_0$ variation consistency with the source voice. 10 testing speakers are randomly selected as the source speakers, and the remaining 10 testing speakers are treated as target speakers, which leads to 100 conversion pairs where all source utterances are used for conversion. The results for different methods are shown in Table \ref{asr-pcc}, where the results for source speech are also reported as the performance upper bound. It can be seen that VQMIVC achieves the lowest CER and WER among all methods, which shows the robustness of the proposed VQMIVC method to preserve the source linguistic content. Besides, we observe that ASR performance drops significantly without using MI (w/o MI), as the converted voice is contaminated by undesired content information entangled with speaker representations. In addition, by providing source pitch representations, we can explicitly and effectively control intonation variations of the converted voice to achieve high $\textit{F}_0$ variation consistency, as indicated by largest $\textit{F}_0$-PCC of 0.781 obtained by the proposed methods.

\begin{figure}[t]
  \centering
  \includegraphics[width=7.5cm]{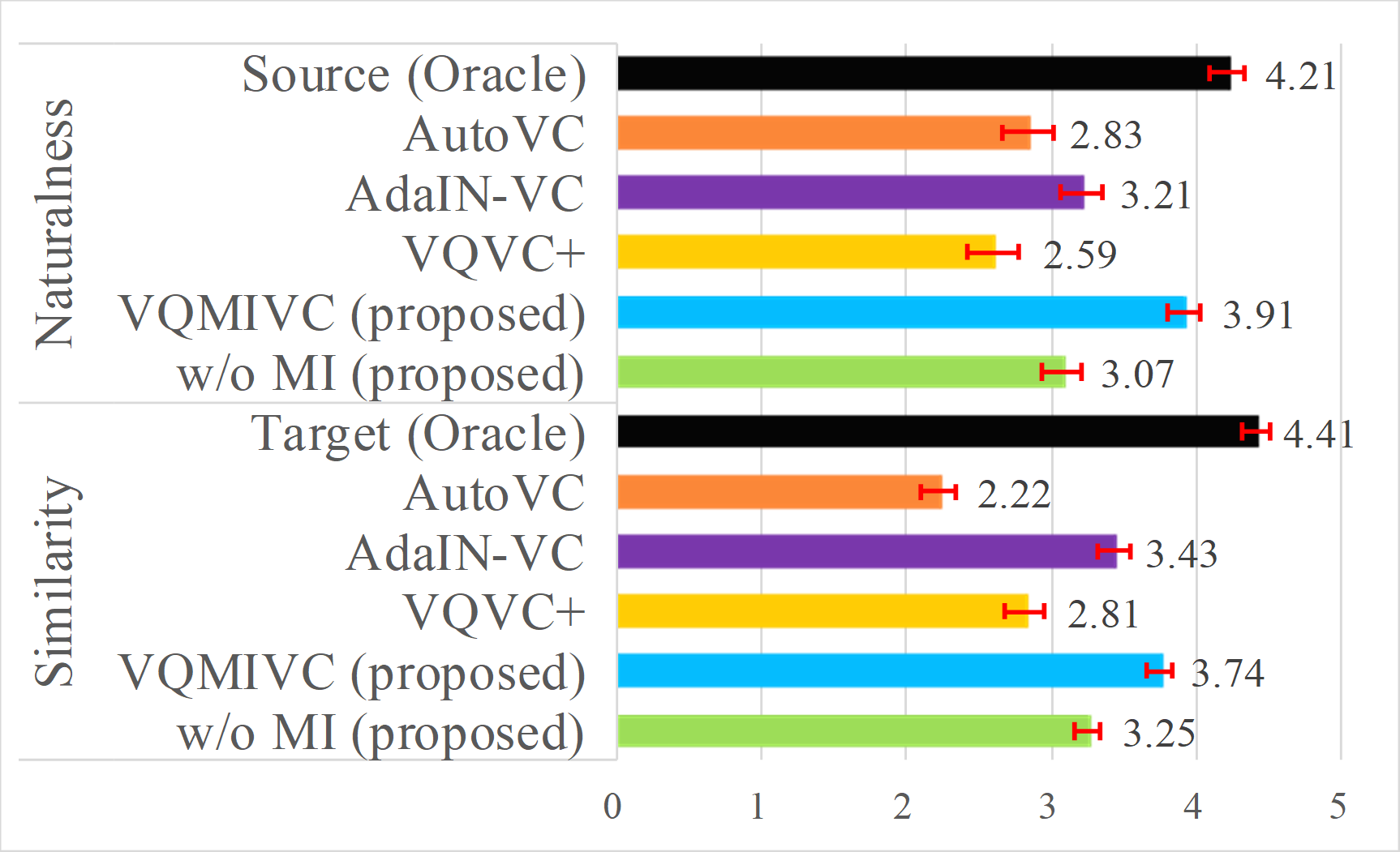}
  \vspace{-0.9em}
  \caption{Comparison results of MOS with 95\% confidence intervals for speech naturalness and speaker similarity.}
  \label{MOS}
  \vspace{-1.7em}
\end{figure}

\vspace{-0.5em}
\subsubsection{Speech naturalness and speaker similarity}
Subjective tests are conducted by 15 subjects to evaluate the speech naturalness and speaker similarity, in terms of 5-point mean opinion score (MOS), i.e, 1-bad, 2-poor, 3-fair, 4-good, 5-excellent. We randomly select two source speakers and two target speakers from the testing speakers, each source or target set contains one male and one female speaker, which results in 4 conversion pairs, where 18 converted utterances from each pair are evaluated by each subject. The scores are averaged across all pairs and reported in Figure \ref{MOS}. Source (Oracle) and Target (Oracle) denote that the speech is synthesized with ground-truth mel-spectrograms of source and target utterances by Parallel WaveGAN respectively. We observe that the proposed method, denoted as `w/o MI', outperforms AutoVC and VQVC+, but is inferior to AdaIN-VC. Pronunciation errors are frequently detected in the converted voice by `w/o MI' via our official listening tests, which can be reflected by the high CER/WER of `w/o MI' in Table \ref{asr-pcc}. These issues can be greatly alleviated by the proposed MI minimization, where improved speech naturalness and speaker similarity are achieved. This indicates that MI minimization facilitates proper SRD to derive accurate content representation and effective speaker representation, which can be used to generate the natural speech with high voice similarity to the target speaker.

\section{Conclusions}

We propose a novel approach by combining VQCPC and MI for unsupervised SRD-based one-shot VC. To achieve proper disentanglement of content, speaker and pitch representations, VC model is not only trained to minimize the reconstruction loss, but also VQCPC loss to explore local structures of speech for content, and MI loss to reduce the correlation between different speech representations. Experiments verify the efficacy of proposed methods to mitigate information leakage issues by learning accurate content representation to preserve source linguistic content, speaker representation to capture desired speaker characteristics, and pitch representation to retain source intonation variations, which results in high-quality converted voice.

\section{Acknowledgements}
This research is partially supported by a grant from the HKSARG Research Grants Council General Research Fund (Project Reference No. 14208718).


% \begin{thebibliography}{9}
% \bibitem[1]{Davis80-COP}
%   S.\ B.\ Davis and P.\ Mermelstein,
%   ``Comparison of parametric representation for monosyllabic word recognition in continuously spoken sentences,''
%   \textit{IEEE Transactions on Acoustics, Speech and Signal Processing}, vol.~28, no.~4, pp.~357--366, 1980.
% \bibitem[2]{Rabiner89-ATO}
%   L.\ R.\ Rabiner,
%   ``A tutorial on hidden Markov models and selected applications in speech recognition,''
%   \textit{Proceedings of the IEEE}, vol.~77, no.~2, pp.~257-286, 1989.
% \bibitem[3]{Hastie09-TEO}
%   T.\ Hastie, R.\ Tibshirani, and J.\ Friedman,
%   \textit{The Elements of Statistical Learning -- Data Mining, Inference, and Prediction}.
%   New York: Springer, 2009.
% \bibitem[4]{YourName17-XXX}
%   F.\ Lastname1, F.\ Lastname2, and F.\ Lastname3,
%   ``Title of your INTERSPEECH 2020 publication,''
%   in \textit{Interspeech 2020 -- 20\textsuperscript{th} Annual Conference of the International Speech Communication Association, September 15-19, Graz, Austria, Proceedings, Proceedings}, 2020, pp.~100--104.
% \end{thebibliography}

\end{document}